\documentclass[11pt]{article}
\usepackage{graphicx,amsmath,amsfonts,color,multicol,multirow}
\def \beq  {\begin{equation}}
\def \eeq  {\end{equation}}
\def \beqar {\begin{eqnarray}}
\def \eeqar {\end{eqnarray}}
\def\sqr#1#2{{\vcenter{\vbox{\hrule height.#2pt
\hbox{\vrule width.#2pt height#1pt \kern#1pt
\vrule width.#2pt}\hrule height.#2pt}}}}

\def\S {{\cal S}}
\def\la {{\langle}}
\def\ra {{\rangle}}
\def\vx {{\vec x}}
\def\vy {{\vec y}}

\def\vf {{\varphi}}

\def\dag {{\dagger}}

\def\Tr {{\rm Tr}}

\def\bp {\bar p}

\def\bD {\bar{D}}
\def\bA {\bar{A}}

\def\bx {\bar{x}}
\def\by {\bar{y}}

\def\bv {\bar{v}}

\def\vx {{\vec x}}

\def\vy{\vec{y}}
\def\vv {\vec{v}}
\def\vu {\vec{u}}

\def\dag {\dagger}
\def\del {\partial}
\def\bdel{\bar{\partial}}

\def\e {\epsilon}
\def\d {\delta}
\def\s {\sigma}

\def\D {{\cal D}}

\def\H {{\cal H}}

\def\G {{\cal G}}

\def\vf {{\varphi}}

\def \H {{\cal H}}

\def\half{\textstyle{1\over 2}}

\begin{document}
\fontfamily{cmr}\fontsize{11pt}{17.2pt}\selectfont
\def \CMP {{Commun. Math. Phys.}}
\def \PRL {{Phys. Rev. Lett.}}
\def \PL {{Phys. Lett.}}
\def \NPBProc {{Nucl. Phys. B (Proc. Suppl.)}}
\def \NP {{Nucl. Phys.}}
\def \RMP {{Rev. Mod. Phys.}}
\def \JGP {{J. Geom. Phys.}}
\def \CQG {{Class. Quant. Grav.}}
\def \MPL {{Mod. Phys. Lett.}}
\def \IJMP {{ Int. J. Mod. Phys.}}
\def \JHEP {{JHEP}}
\def \PR {{Phys. Rev.}}
\def \JMP {{J. Math. Phys.}}
\def \GRG{{Gen. Rel. Grav.}}
\begin{titlepage}
\null\vspace{-62pt} \pagestyle{empty}
\begin{center}
\rightline{CCNY-HEP-07/5}
\rightline{May 2007}
\vspace{1truein} {\Large\bfseries
The robustness of the vacuum wave function and other}\\
\vskip .1in
{\Large\bfseries  matters for Yang-Mills theory}\\
\vspace{.5in}
{\large DIMITRA KARABALI}\\
\vspace{.1in}{\itshape Department of Physics and Astronomy\\
Lehman College of the CUNY\\
Bronx, NY 10468}\\
E-mail: \fontfamily{cmtt}\fontsize{11pt}{15pt}\selectfont dimitra.karabali@lehman.cuny.edu\\
\fontfamily{cmr}\fontsize{11pt}{15pt}\selectfont
\vspace{.4in}
{\large V. P. NAIR}\\
\vspace{.1in}{\itshape Physics Department\\
City College of the CUNY\\  
New York, NY 10031}\\
E-mail:
{\fontfamily{cmtt}\fontsize{11pt}{15pt}\selectfont vpn@sci.ccny.cuny.edu}

\vspace{.4in}
\centerline{\large\bf Abstract}
\end{center}
In the first part of this paper, we present a set of
simple arguments to show that the two-dimensional gauge anomaly and the $(2+1)$-dimensional Lorentz symmetry determine the leading Gaussian term in the vacuum wave function of $(2+1)$-dimensional Yang-Mills theory. This is to highlight the robustness of the wave function and its relative insensitivity to the choice of regularizations.
We then comment on the correspondence with the explicit calculations done in earlier papers.
We also make some comments on the nature of the gauge-invariant configuration space
for Euclidean three-dimensional gauge fields (relevant to $(3+1)$-dimensional Yang-Mills theory).
\end{titlepage}
\pagestyle{plain} \setcounter{page}{2}

\section{Introduction}
There has recently been a revival of interest in the Hamiltonian approach to
Yang-Mills theories in $2+1$ and in $3+1$ dimensions. This is partly because of
earlier work where it was noticed that in a Hamiltonian approach in $2+1$ dimensions,
one could utilize some of the niceties of two-dimensional gauge theories \cite{KN, KKN1, KKN2}.
In particular, one could choose the $A_0 =0$ gauge and for the remaining
two spatial components a matrix parametrization of the form
$A= \half (A_1 +i A_2) = - \del M M^{-1}$, where $M$ is a complex matrix, 
could be used. On the matrix $M$, gauge transformations act homogeneously
by left-multiplication and hence the reduction to the gauge-invariant set of variables
is more easily accomplished.
This led to the computation of the volume element for
the gauge-invariant configuration space, the reduction of the Hamiltonian
(to gauge-invariant variables)
and the computation of the vacuum wave function.
The expectation value of the Wilson loop could be calculated and gave a value for
string tension in good agreement with lattice simulations.

There have been more recent attempts to extend this analysis to obtain estimates of
glueball masses \cite{minic}. There have also been attempts to extend the
discussion of the gauge-invariant configuration space to $3+1$ dimensions,
where results have been more limited \cite{NY, Freidel}.
It is also worth mentioning that there have been a number of other analyses
which are similar in spirit, i.e., within the general framework of the Hamiltonian approach
to Yang-Mills theory, but different in details \cite{others}.

The calculations presented in \cite{KN, KKN1, KKN2} are simplified by the 
parametrization we used and known results for
two-dimensional gauge fields. Nevertheless, they are still quite involved. In particular,
we need to have proper regularization for all the terms in the
Hamiltonian, the wave function, etc. While this was sorted out in detailed calculations,
the reason why each component-result in the chain of argument should be true
was not always transparent. Can we understand the essential elements of these
results based on simple invariance arguments so that sensitivity to
regularization is clearly eliminated? The following comments will address this question.
We will present arguments to show that the leading Gaussian term in
the wave function as calculated in \cite{KKN1, KKN2} is obtained 
from the two-dimensional gauge anomaly and $(2+1)$-dimensional 
Lorentz invariance. Detailed properties of regularization are not needed.
We will then comment on the points of correspondence between these
arguments and the detailed calculations of the earlier papers.
In the last section, we present some considerations
 on the gauge-invariant configuration space
of three-dimensional Euclidean gauge fields which is relevant for a Hamiltonian
analysis of $(3+1)$-dimensional gauge theories. 

\section{Robustness of the wave function}

We will start with a sequence of arguments which will show that the leading terms in the wave function have a certain degree of robustness.
For this we will use the two-dimensional anomaly calculation combined with $(2+1)$-dimensional Lorentz (Galilean) invariance and, to some extent, the perturbative limit.

\vskip .1in\noindent{$\underline{The ~volume ~element~ for ~gauge-invariant ~configurations}$}

We start with the calculation of the volume element
on the gauge-invariant configuration space. Once we have chosen the gauge condition
$A_0 =0$, the spatial components of the gauge potential may be parametrized as
\beq
A= - \del M M^{-1}, \hskip .4in \bA = M^{\dagger -1} \bdel M^\dagger
\label{c0}
\eeq
Here $M$ is a complex matrix which is an element of the complexification of the gauge group.
Thus, for the group $SU(N)$ which we shall consider here, $M \in SL(N, \mathbb{C})$.
The gauge-invariant hermitian matrix $H= M^\dagger M$ will describe the physical
(gauge-invariant) degrees of freedom. It may be considered as parametrizing
$SL(N, {\mathbb C}) / SU(N)$. (A basis for the Lie agebra of $SU(N)$, in the
fundamental representation, will be taken as the set of $N\times N$ traceless hermitian
matrices $t^a$, $a= 1, 2, \cdots,  N^2-1$, with $[t^a, t^b]= i f^{abc} t^c$ and
$\Tr (t^a t^b) = \half \delta^{ab}$.)

Denoting the space of gauge potentials $\{ A, \bA \}$
as ${\cal A}$ and the set of all gauge transformations as ${\cal G}_*$, we are interested
in the volume element of the gauge-invariant configuration space ${\cal A}/ {\cal G}_*$.
The parametrization (\ref{c0}) leads to
\beq
d\mu ({\cal A}/ {\cal G}_*) = \det (-D \bD ) ~ d\mu (H)
\label{c1}
\eeq
where
$d\mu (H)$ is the Haar measure on the coset space $SL(N, {\mathbb C}) / SU(N)$.
The determinant in this equation can be calculated by evaluating its variation.
Defining
$\Gamma = \log \det (-D\bD )$, we can write
\beq
{\delta ~\Gamma \over \delta \bA^a(\vx )}~= -i~\Tr\Bigl[ \bD^{-1}(\vx,\vy) 
T^a\Bigr]_{\vy\rightarrow \vx} \label{c2}
\eeq
Here $(T^a)_{mn}=-if^a_{mn}$ are the generators  of the Lie algebra in the adjoint
representation. The coincident-point limit of the Green's function
$\bD^{-1}(\vx,\vy)$ is
singular and needs regularization. Since the volume element $d\mu ({\cal A}/ {\cal G}_*)$
must be gauge-invariant, we choose
a gauge-invariant regularization.
For any gauge-invariant regularization, this leads to
\beq
\Tr \left[ \bD^{-1}_{reg}(\vx,\vy) T^a \right]_{\vy\rightarrow \vx}~= {2c_A \over \pi}~
\Tr \left[ \Bigl(A(\vx ) -M^{\dag -1}(\vx ) \partial M^\dag (\vx) \Bigr)t^a\right] \label{c3}
\eeq
where $c_A$ is the quadratic Casimir invariant for the adjoint representation
defined by
$f^{amn} f^{bmn} = c_A \delta^{ab}$.
Using this result in (\ref{c2}), and with a similar result for the variation of 
$\Gamma$ with respect to $A^a$, and integrating, we get,
up to an additive constant, $\Gamma = 2c_A \S_{wzw} (H)$, where
$\S_{wzw}(H)$ is the Wess-Zumino-Witten (WZW) action for the hermitian matrix field
$H$,
\beq
\S_{wzw}(H) = {1 \over {2 \pi}} \int \Tr (\partial H \bar{\partial} H^{-1}) +{i
\over {12 \pi}} \int \epsilon ^{\mu \nu \alpha} \Tr ( H^{-1} \partial _{\mu} H H^{-1}
\partial _{\nu}H H^{-1} \partial _{\alpha}H)
\label{c4}
\eeq
For the volume element (\ref{c1}), we then have, up to a multiplicative
constant,
\beq
d\mu ({\cal A}/ {\cal G}_*) =  d\mu (H)~\exp \left( 2c_A \S_{wzw}(H) \right)
\label{c5}
\eeq
The calculation in (\ref{c3}) is essentially the calculation of the gauge anomaly in two dimensions and,
therefore, the result (\ref{c5}) is quite robust; different regulators
will lead to the same result so long as gauge invariance is preserved. 
\vskip .1in\noindent{$\underline{The ~action ~of ~T~ on ~J^a}$}

This result is closely related to another, namely, the action of the kinetic energy operator
on the current 
\beq
J^a = {c_A \over \pi }~ (\del H H^{-1})^a\label{c5a}
\eeq
This is the current for the WZW action 
in (\ref{c4}). The current $J^a$ is the gauge-invariant variable in terms of which all observables can be constructed. For the action of $T$, we find
\beqar
 T~J^a(\vx) &=& \int d^2y~ {E^2 \over 2e^2}~ J^a (\vx )\nonumber\\
 &=& -{e^2\over 2} \int d^2y {\delta^2 J^a(\vx)\over 
\delta \bA^b(\vy) \delta A^b(\vy)} ~={e^2c_A\over 2\pi} M^{\dag am} \Tr \left[ T^m
\bD ^{-1}(\vy,\vx) 
\right]_{\vy \rightarrow \vx}\nonumber\\
 &=& m ~J^a(\vx) \label{c6}
 \eeqar
where $m = e^2 c_A /2\pi$. Notice that the basic calculation involved is the same as in
(\ref{c3}); therefore, this result also follows from the two-dimensional gauge anomaly.

There should be no surprise that the two results (\ref{c5}) and (\ref{c6}) are related.
As argued in \cite{KKN1}, the self-adjointness of the kinetic energy operator
$T$ relates it to the gauge-invariant volume element.

\vskip .1in\noindent{$\underline{Identifying~ the ~vacuum~ wave function}$}

Consider now the vacuum wave function which we may write as $\Psi_0 = e^P$
where $P$ is a functional of the current $J$ and its derivatives.
We write $P = -\beta V +\cdots$, where $V$ is the potential energy $\int B^2 /2e^2$, or
$(\pi /m c_A )\int \bdel J^a \bdel J^a $ in terms of the current.
(These have to be understood with proper regularization; we will not need the explicit form of the regularization for the argument we present. It is discussed in the next section.)
The action of the kinetic energy operator on $V$, considered as a functional of $J$,
leads to an equation of the form
\beq
[T, V] = a ~V ~+ {4\pi \over c_A} \int  (\D \bdel J)^a ~{\delta \over \delta J^a}
\label{c7}
\eeq
where 
\beq
{\cal{D}}_{x~ab} = {c_A\over \pi}\partial_x \delta_{ab} +if_{abc}J_c (\vx)
\label{c8}
\eeq
Notice that, on dimensional grounds, $\int (\delta^2 V/\delta \bA \delta A)$
should be proportional to $V$. This is the reason for postulating the first term 
on the right hand side in
(\ref{c7}). The computation of the coefficient $a$ has to be done with proper 
regularization.
However, the second term does not involve the intricacies of regularization, it follows directly from the variation of $\int B^2$ with respect to $A$.

Using (\ref{c7}), we find for the action of the Hamiltonian on $\Psi_0 \approx e^{-\beta V}$, \beq
\H ~\Psi_0 = (T+V) \Psi_0 = e^P ~( V - \beta a V +\cdots )
\label{c9}
\eeq
where the omitted terms involve derivatives (or momenta $k$) due to the second set of terms
in (\ref{c7}). In an expansion in powers of $k/e^2$, these are negligible.
Thus, to lowest order in $k/e^2$, we must cancel the $V$-dependent terms to get
a solution to the vacuum wave function. This requires $\beta = 1/a$. The vacuum wave function, to this order, is thus
\beq
\Psi_0 \approx \exp (- V/a )
\label{c10}
\eeq

We now go back to the result (\ref{c6}). This states that, in the extreme strong coupling limit
where we neglect $V$ entirely, $J^a$ is an eigenstate of $T$ with eigenvalue
$m$. Notice that we can write this state as $J^a \Psi_0$ since
$\Psi_0 \approx 1$ in the extreme strong coupling limit.
We can see that, once we include the modification to $\Psi_0$ due to $V$, this is the corrected eigenstate of the Hamiltonian to first order in $V$ and in $k/e^2$. In fact, we find
\beqar
(T+V ) ~J^a \Psi_0 &=&  e^P \left( T+V - \beta [T,V] + \cdots \right) J^a\nonumber\\
&=& \left( m + {k^2 \over a} +\cdots \right) ~ J^a e^P ~+
e^P J^a (V - \beta a V +\cdots )\nonumber\\
&=& \left( m + {k^2\over a} +\cdots \right) ~J^a \Psi_0
\label{c11}
\eeqar
We see that we have, indeed, found the corrected eigenstate to first order in the
$1/e^2$ expansion; the eigenvalue is $m + k^2/a$.
This eigenvalue must have the form
$m + k^2/2m$ for this to become the standard relativistic formula for the energy, to this order.
This identifies $a$ as $2m$. Going back to (\ref{c7}), we can now write
\beq
[T, V] = 2m~V ~+ {4\pi \over c_A} \int  (\D \bdel J)^a ~{\delta \over \delta J^a}
\label{c12}
\eeq
Notice that we have only assumed $a$ to be nonzero. Its actual value is then fixed by Lorentz
invariance and the action of $T$ on $J^a$. Since the latter is given by
the anomaly, and hence is quite robust, we see that (\ref{c12}) is unambiguously obtained.
The vacuum wave function to this order of calculation is thus
$\Psi_0 \approx \exp ( - V /2m )$. (In (\ref{c11}), we have only used the first correction
to $m$ in a $k/m$-expansion. As shown elsewhere \cite{KKN2}, there is a set of terms
which add up to give the full relativistic expression for the energy.)

Starting with this formula for the vacuum wave function, in reference \cite{KKN2}, we obtained
a series for $P$,  in powers of $k/m$. The leading terms, with two powers of the current $J$, were
summed up to give
\beq
\Psi_0 \approx \exp \left[ - {{2 \pi ^2} \over { e^2 {c_A}^2}} \int  \bdel J_a \left[ { 1 \over {\bigl( m
+ \sqrt{m^2
-\nabla^2 } \bigr)}} \right] \bdel J_a + {\cal O}(J^3) \right]
\label{c13}
\eeq
So far, we have basically argued for the robustness of the leading term of this expression
where we neglect the momenta or $\nabla^2$. 
(It is worth noting that this is also the form which gives the
fully relativistic formula $\sqrt{k^2 +m^2}$ for the action of $T+V$
on $J^a \Psi_0$.)

\vskip .1in\noindent{$\underline{Another ~argument ~for ~the ~form~ of~ \Psi_0}$}

There is another check of this formula that we can do, starting from
(\ref{c5}). Using the formula for the gauge-invariant volume element, we can write for the inner product of the wave functions,
\beqar
\la 1\vert 2\ra &=& \int d\mu ({\cal A}/{\cal G}_*) ~ \Psi_1^* \Psi_2\nonumber\\
&=& \int d\mu (H) e^{2c_A \S_{wzw}(H)}~ \Psi_1^* \Psi_2
\label{c14}
\eeqar
As we have argued elsewhere \cite{KN, KKN1}, the WZW action in the exponent for the volume element
is related to a mass gap. This is seen explicitly by writing $\Psi = \exp[ - c_A \S_{wzw}(H) ]
~\Phi$. The inner product then simplifies as
\beq
\la 1 \vert 2\ra = \int d\mu (H)~ \Phi_1^* \Phi_2
\label{c15}
\eeq
The Hamiltonian acting on $\Phi$'s is given by $\H_\Phi = e^{c_A \S_{wzw}}
\H e^{-c_A \S_{wzw}}$. For the argument we are going to present, it is sufficient to
consider the small $\vf$-expansion where $H = \exp( t^a \vf^a )
\approx 1 + t^a \vf^a$. In this case
\beqar
c_A \S_{wzw} &\approx& - {c_A \over 4\pi} \int \del \vf^a \bdel \vf^a + \cdots\nonumber\\
\H_\Phi &\approx& {1\over 2} \int \left[ -{\delta \over \delta \phi^a \delta \phi^a }
+ \phi^a ( m^2 -\nabla^2 ) \phi^a \right] ~+\cdots
\label{c16}
\eeqar
where $\phi^a = \sqrt{ c_A (-\nabla^2) /8\pi m} ~\vf^a$. We see that the leading term in
$\H_\Phi$ corresponds to a free field of mass $m$ (actually $dim G$ fields, counting the multiplicity
due to the index $a$.) To arrive at this result we have used the fact that
\beq
T \approx m \left[ \int \vf^a {\delta \over \delta \vf^a} -
{4\pi \over c_A} \int {\delta \over \delta \vf^a (x)} \left( {1\over -\nabla^2} \right)_{x,y}
{\delta \over \delta \vf^a (y)}  +\cdots \right]
\label{c17}
\eeq
The first term in this expression follows from (\ref{c6}). The second term does not involve the intricacies of regularization; it is just the rewriting of $- \delta^2 /\delta A^2$ to the perturbative linear order
in $\vf$. (If we write $A \approx -\del \theta $, $\vf$ is given as $\vf = \theta +{\bar \theta}$, and we get
the second term on the right hand side of (\ref{c17}) when $\delta /\delta A~\delta/\delta \bA$ acts
on functionals of $\vf$.) Thus, to the order we have calculated, (\ref{c17}) also follows from the gauge anomaly calculation.

Since (\ref{c16}) is the Hamiltonian for free fields, the vacuum wave function is trivially
constructed as
\beq
\Phi_0 \approx \exp \left[ - {1\over 2} \int \phi^a \sqrt{ m^2 - \nabla^2} ~\phi^a \right]
\label{c18}
\eeq
Going back to $\Psi_0$, we find
\beqar
\Psi_0 &=& e^{-c_A \S_{wzw}}~ \Phi_0\nonumber\\
&\approx& \exp \left( {c_A \over 4\pi} \int \del \vf^a \bdel \vf^a +\cdots \right)~ 
\exp\left[ - {c_A \over 16\pi m}\int (-\nabla^2 \vf)^a \sqrt{m^2 -\nabla^2}~ \vf^a +\cdots \right]
\nonumber\\
&\approx& \exp \left[ - {c_A \over \pi m} \int (\bdel \del \vf^a) \left[{1\over m + \sqrt{m^2 - \nabla^2}}\right]
(\bdel \del \vf^a) + \cdots \right]\label{c19}
\eeqar

The basic argument can now be formulated as follows. Let us say we start with the Yang-Mills theory in $2+1$ dimensions. Then
the inner product is given by (\ref{c14}); further $\Psi_0$ should be a functional of
$J$. So far we do not need to make any small $\vf$-approximations.
Now we can say that, whatever $\Psi_0$ is, it should agree with (\ref{c19}) 
in the small $\vf$-limit. The only functional of
$J$ which has this property is (\ref{c13}). (It is easily checked that (\ref{c13}) agrees with
(\ref{c19}) in the small $\vf$-limit, using $J = (c_A/\pi) \del H H^{-1} \approx (c_A/\pi ) \del \vf$.)
Thus, we see that, in short, the volume element and the perturbative small $\vf$-limit
restrict $\Psi_0$ to the form (\ref{c13}). The formula for the measure,
which is determined by the anomaly, and the form of $T$ in (\ref{c17}),
which is also determined by the anomaly, are the key ingredients for this argument.

\vskip .1in\noindent{$\underline{How ~does~ this ~apply~ to ~the~ string ~tension?}$}

The vacuum expectation value of any operator ${\cal O}$
is given by
\beq
\la {\cal O} \ra = \int d\mu ({\cal A} /{\cal G}_*)~\Psi_0^* \Psi_0 ~ {\cal O}
= \int d\mu ({\cal A} /{\cal G}_*)~e^{-S} ~ {\cal O}
\label{c20}
\eeq
where $S$ is defined by $\Psi^*_0 \Psi_0 = e^{-S}$. The expectation value is, thus, the functional average in a two-dimensional
gauge theory with the action $S$.
Based on arguments given above, for modes of low momentum, the wave function for the vacuum can be taken
as
\beq
\Psi_0 \approx \exp \left[ - {\pi \over 2m^2 c_A}\int  \bdel J^a \bdel J^a\right] =
\exp\left[ - {1\over 8g^2 }\int F^a_{ij} F^a_{ij}\right]
\label{c21}
\eeq
where $g^2 = me^2$, so that $S \approx  S^{(2)}_{YM}$, where $S^{(2)}_{YM}$ is the two-dimensional Yang-Mills action with coupling constant $g^2$. The 
expectation value of the Wilson loop operator (in the representation $R$) then obeys an area law given by
\beqar
\la W_R(C, A) \ra &=& \int  d\mu ({\cal A}/{\cal G}_*) e^{- S} ~W_R(C, A)\nonumber\\
&\approx& \int  d\mu ({\cal A}/{\cal G}_*) e^{- S^{(2)}_{YM}} ~W_R(C, A)
\sim \exp \left[ - \sigma_R {\cal A}(C)\right]
\label{c22}
\eeqar
where ${\cal A}(C)$ is the area of the loop $C$ and the string tension
$\sigma_R$ is given by
\beq
\sigma_R = e^4~ {c_A ~c_R \over 4\pi}
\label{c23}
\eeq
As mentioned elsewhere, and as the following table shows,
this formula is in good agreement with the lattice estimates \cite{teper}, the difference being less than $3\%$ for all cases, and less than $0.88\%$ as
$N\rightarrow \infty$, even though
the deviations are
still statistically significant \cite{bringoltz}.

\begin{table}[!t]
\begin{center}
\begin{tabular}{| p{1.3cm}| p{1.7cm} p{1.5cm} p{1.5cm} p{1.5cm} p{1.5cm} p{1.5cm}| }
\hline\hline
Group&\multicolumn{6}{c|}{Representations}\\
\hline
&k=1& k=2 & k=3 & k=2 &k=3 &k=3\\
&Fund.&antisym&antisym&sym&sym&mixed\\
\hline\hline
$SU(2)$&0.345&&&&&\\
&{\color{red}0.335}&&&&&\\
\hline
$SU(3)$&0.564&&&&&\\
&{\color{red}0.553}&&&&&\\
\hline
$SU(4)$&0.772&0.891&&1.196&&\\
&{\color{red}0.759}&{\color{red}0.883}&&{\color{red}1.110}&&\\
\hline
$SU(5)$&0.977&&&&&\\
&{\color{red}0.966}&&&&&\\
\hline
$SU(6)$&1.180&1.493&1.583&1.784&2.318&1.985\\
&{\color{red}1.167}&{\color{red}1.484}&{\color{red}1.569}&{\color{red}1.727}&{\color{red}2.251}&{\color{red}1.921}\\
\hline
$SU(N)$&0.1995 $N$&&&&&\\
$N\!\!\rightarrow \!\!\infty$&\color{red}0.1976 $N$&&&&&\\
\hline\hline
\end{tabular}\\
\vskip .2in
Comparison of $\sqrt{\sigma}/e^2$ as predicted by (\ref{c23}) (upper entry) and lattice estimates
(lower entry, in red) from \cite{teper, bringoltz}. $k$ is the rank of the representation.
\end{center}
\end{table}

We have argued that the leading term of the vacuum wave function (\ref{c13}),
and hence the leading term in $S$ (which is quadratic in the currents),  is quite robust.
Therefore, if there are any corrections to the string tension, they should arise, not from
modification of the wave function, but due to the approximation of $S$ by
$S^{(2)}_{YM}$ in the evaluation of the expectation value
(\ref{c22}). Thus corrections to $\sigma$ should be due to terms in $S$
which are higher than quadratic in the $J$'s.

On general grounds, we should expect
some corrections to the formula for the string tension.
It has been argued that the ratios of string tensions should deviate from
the ratios of Casimir invariants on the basis of the $1/N$-expansion
\cite{armoni}. 
Also, for Wilson loops in the adjoint representation (or other representations
which are invariant under the center of the group), we should expect screening rather than confinement or area law. We have presented reasons to show how
screening and the corresponding string-breaking effect can arise from a judicious resummation of the higher order corrections which can lead to the formation of 
color-singlet bound states of a ``gluon" with the external charge whose world line trajectory is represented by (part of) the Wilson loop. An estimate of the string-breaking energy along these lines gives a result within $8.8\%$ of the lattice estimates \cite{AKN}.

\section{Correspondence with explicit calculations}

\vskip .1in\noindent{$\underline{How ~do ~we ~regularize~ the~ Hamiltonian?}$}

We now turn to the question: How are the results given so far explicitly realized when we solve the Schr\"odinger equation after regularization of the Hamiltonian?
This was done in some detail in \cite{KKN1}, so the following comments are more in the nature of clarifying remarks. The Hamiltonian consists of the kinetic term $T$, which is a functional differential operator, and $V$, the potential energy.
Since Lorentz transformations can mix the two, there has to be a concordance between the regularization of these two terms to ensure that the full theory has Lorentz symmetry.

In the regularized expression for any quantity in field theory, one can have terms which are suppressed by powers of $k/M$ where $k$ is a typical momentum and $M$ is the regulator mass. The details of such terms differ from regulator to regulator and constitute regularization ambiguities. These regularization-dependent terms are, of course, negligible if we consider processes of momenta $k\ll M$. In other words, once we introduce a regulator, we must apply the results only to processes with $k\ll M$. This is well-known lore in field theory, but is worth emphasizing in the context of regularization of terms in the Hamiltonian.
Now, of the two terms in the Hamiltonian, the kinetic energy requires more care regarding regularization, so we consider it first.
As a regularized expression, we may take the kinetic energy operator as
\beqar
T_{(\e )}
&=&{e^2 \over 2} \int_{u,v} \Pi_{rs} (\vu,\vv)
\bp_r (\vu) p_s (\vv)\label{reg1} \\
\Pi_{rs} (\vu,\vv) &=& \int_x \bar{\G} _{ar} (\vx,\vu) K_{ab}(\vx) \G _{bs} (\vx,\vv) \nonumber
\eeqar
where $K_{ab} = 2 \Tr (t_a H t_b H^{-1})$ is the adjoint representative of
$H$. The functions $\bar{\G} _{ma} (\vx,\vy)$, $\G _{ma} (\vx,\vy)$ are given by
\beqar
\bar{\G} _{ma} (\vx,\vy)  &=& {1\over \pi (x-y)}   \Bigl[ \d _{ma} - e^{-|\vx-\vy|^2/\e} \bigl(
K(x,\by) K^{-1} (y, \by) \bigr) _{ma}\Bigr] \nonumber\\
\G _{ma} (\vx,\vy)  &=&  {1\over \pi (\bx - \by )} \Bigl[ \d _{ma} - e^{-|\vx-\vy|^2/\e} \bigl(
K^{-1}(y,\bx) K (y, \by) \bigr) _{ma}\Bigr]
\label{reg2}
\eeqar
These are the regularized versions of the corresponding Green's functions
\beq
\bar{G} (\vec{x},\vec{y}) = {1 \over {\pi (x-y)}} ~, ~~~~      G (\vec{x},\vec{y}) = {1 \over {\pi (\bar{x}-\bar{y})}} \label{reg3}
\eeq
The parameter $\sqrt{\e}$ acts as a short-distance cut-off; it is the regularization parameter, taken to be arbitrarily small compared to other distance scales in the theory.
In the naive $\e \rightarrow 0$ limit, we find
\beq
T_{(\e )}\biggr]_{(\e \rightarrow 0 )}=
{ e^2 \over 2} \int d^2x~ E^2  =
-\frac{e^2}{2}\int \frac{\delta ^2}{\delta {A}^a\delta \bar{{A}}^a} \label{reg4}
\eeq
so that (\ref{reg1}) can indeed be interpreted as the regularized version
of the kinetic energy.

One can now consider the action of this operator on functionals $\Psi(\lambda' )$,
which is some product of fields and their derivatives with
an average separation of points between fields being $\sqrt{\lambda'}$.
When $T_{(\e )}$ acts on this, it can generate terms which diverge as $\e \rightarrow 0$,
terms which are finite as $\e \rightarrow 0$ and terms which vanish as $\e \rightarrow 0$.
The first type of terms would indicate that we must do an additional subtraction to define a `renormalized' kinetic energy operator. The second set of terms corresponds to physically meaningful results. The last set of terms represents regularization ambiguities. They vanish
when $\e$ goes to zero, but they may be in the form of powers of $\e /\lambda'$.
If we take $\lambda'$ comparable to $\e$, the results can be ambiguous.
(For example, a different regularization may give different results for these terms.)
The correct procedure is to keep $\e$ much smaller than $\lambda'$; the regularization
in (\ref{reg1}, \ref{reg2}) only applies with this caveat.

The regularized expression for the potential energy can be taken as
\beqar
V_{(\lambda' )}&=& {\pi \over {m c_A}} \biggl[ \int_{x,y} \s (\vx,\vy;\lambda' ) \bdel J_a (\vx) (K(x,\by)
K^{-1} (y,\by))_{ab} \bdel J_b (\vy) - {{c_A {\rm dim} G} \over {\pi^2
\lambda'^{2}}} \biggr] \nonumber\\
\sigma (\vec{x}, \vec{y} ; \lambda') & = &{1\over \pi \lambda'} {\exp\left[{{-|\vec{x}-\vec{y}|^2/ \lambda' }}~\right]}
\label{reg5}
\eeqar
In using this expression for solving the Schr\"odinger equation, we will encounter
terms like $[T_{(\e )}, V_{(\lambda' )}]$, in other words, the action of
$T$ on $V$. From what was stated earlier, for consistency, we must keep
$\lambda'$ much larger than $\e$. Explicit calculation then shows that
\beq
T_{(\e )} ~V_{(\lambda' )} = 2m \left[ 1 + \half \log (\lambda' /2\e )\right] ~V_{(\lambda' )}~+\cdots
\label{reg6}
\eeq
where the omitted terms correspond to powers of $\e$ or $\lambda'$.
This equation shows that we have a potential log-divergence. In addition to the regularization,
we must define a renormalized $T_{(\lambda )}$ as
\beqar
T_{(\lambda )}&=& T_{(\e )} + { e^2 \over 2} \log ({2\e / \lambda })~ {\cal Q}\nonumber\\
 {\cal Q}&=& \e \int \s (\vu,\vv;\e) K_{rs} (u, \bv)~\Bigl( \bp_r (\vu) -i \bdel J_r(\vu )\Bigr)~ p_s(\vv)
 \label{reg7}
\eeqar
$T_{(\lambda )}$ corresponds to a subtraction scale of $\lambda$. Since we are interested in
the ``local" operator $T$, eventually we must take $\lambda$ to be very small compared to
the distance scales in the theory, i.e., $\lambda \ll e^{-4}$.
Using $T_{(\lambda )}$ we find
\beq
T_{(\lambda )} V_{(\lambda' )} = 2m \left[ 1 + \half \log (\lambda' /\lambda )\right] ~V_{(\lambda')} ~+\cdots
\label{reg8}
\eeq
\vskip .1in\noindent{$\underline{Lorentz ~transformation ~once ~more}$}

Consider now an infinitesimal Lorentz transformation corresponding to velocity
$v_i$.
For the electric and magnetic fields we have
\beq
\delta E_i \approx -\epsilon_{ij} v_j B, \hskip .3in \delta B \approx \epsilon_{ij} v_i 
E_j
\label{reg9}
\eeq
For simplicity, consider a transformation along the $x$-axis, so that $v_2 =0$.
The transformation of the Hamiltonian is now given as
\beqar
\delta {\cal H} &=& \delta T_{(\lambda )} ~+~ \delta V_{(\lambda ')}\nonumber\\
&=& v_1 \int (BE_2)_{(\lambda)} + v_1 \int (BE_2 )_{(\lambda')}
\label{reg10}
\eeqar
The two terms on the right hand side must combine to produce twice the momentum
density $P_1 \sim \int BE_2$. Now, for $\int (BE_2)_{(\lambda')}$, there are no modes of
momenta larger than $1/ \sqrt{\lambda'}$, on average. For this to combine with the first term, we must therefore conclude that the smallest value for $\lambda$ must be $\lambda'$.
The consistent regularization, keeping as many modes as possible for both terms would be to have $\lambda = \lambda'$, with $e^2 \ll 1/\sqrt{\lambda}$.
Thus $\H = T_{(\lambda )} + V_{(\lambda )}$, and, going back to
(\ref{reg8}), we get
\beq
T_{(\lambda )} ~V_{(\lambda )} = 2m ~V_{(\lambda )}
\label{reg11}
\eeq
This result holds when $\lambda$ is taken to be very, very small, $\lambda \rightarrow 0$, keeping
$\e \ll \lambda \ll e^{-4}$. This is effectively the result (\ref{c12})
and the construction of the wave function then follows the arguments given after that equation.

Even though the Lorentz transformation properties were not explicitly used
in \cite{KKN1}, the regularization and detailed calculations presented there followed
the same general approach and gave the result (\ref{reg11}). 
It is also worth mentioning that there are regularizations in the literature
which do not lead to (\ref{reg11}), or (\ref{c12}), and which, from our arguments, do not respect the Lorentz symmetry \cite{minic}. (Mansfield in \cite{others} also presents another regularization, and also raises the question of Lorentz invariance.)

\section{The configuration space for $3$-dimensional gauge fields: general comments}

We now turn to some general properties of the gauge-invariant configuration space for Euclidean gauge fields in three spatial dimensions. This would be appropriate for a Hamiltonian analysis for $(3+1)$-dimensional gauge theories in the $A_0=0$ gauge, or
for a covariant path integral calculation for the (Wick-rotated version of) $(2+1)$-dimensional Yang-Mills theory.
\vskip .1in\noindent{$\underline{Is~ the~ volume~ of~ the ~configuration ~space~ finite?}$}

 For two-dimensional gauge fields, the total volume of the configuration space is
 \beq
 \int d\mu ({\cal C}) = \int d\mu (H)~ e^{2c_A \S_{wzw}(H)} ~< \infty
 \label{gen1}
 \eeq
 This is the partition function of the hermitian WZW model and is finite with some regularization (to a finite number of modes). The contrast to  be emphasized here is with the Abelian theory for which $c_A =0$ and the integral diverges for each mode. This result is important for two reasons.
 First of all, it is possible to find configurations which are separated by an infinite distance
 on the configuration space ${\cal C}$. The finiteness of $\int d\mu ({\cal C})$ shows that these have zero transverse measure, i.e., zero volume in the directions transverse to the
line connecting the two configurations.
 Such far-separated configurations are therefore not imporatnt to the question of the spectrum of the Laplacian (i.e., the kinetic energy operator) on ${\cal C}$. Secondly, in continuation of this reasoning, we see that $\S_{wzw}(H)$ provides a cutoff for low momentum modes.
 This property is crucial for the existence of a mass gap.
 
 One can now ask the question whether similar properties are obtained for the three-dimensional gauge fields. There have been a number of attempts at calculations of the volume element for the $(3+1)$-dimensional theory \cite{NY, Freidel}.
 These have generally been in special parametrizations for the fields. However, here, we shall consider some general properties. The naive volume element $[dA]/vol({\cal G}_*)$
 is difficult to analyze, so it is useful to define it as the limit of a ``regularized" version as
 \beq
 d\mu ({\cal C})_{3d} = {[dA]\over vol({\cal G}_*)}~  \exp \left( - {1\over 4\mu}\int F^2 \right)\Biggr]_{\mu \rightarrow \infty}
 \label{gen2}
 \eeq
 where $\mu$ has the dimensions of mass. The right hand side is the functional measure for the Euclidean (Wick-rotated) version of $(2+1)$-dimensional Yang-Mills theory with a coupling constant $e^2 =\mu$. Therefore we can evaluate various quantities by the Hamiltonian techniques we have developed for the $(2+1)$-dimensional theory. In particular, the total volume is given by the Euclidean version of the vacuum-to-vacuum transition amplitude,
 \beqar
 \int d\mu ({\cal C})_{3d} &=& \int {[dA]\over vol({\cal G}_*)}~  \exp \left( - {1\over 4\mu}\int F^2 \right)\Biggr]_{\mu \rightarrow \infty}
 \nonumber\\
 &=& \la 0\vert ~e^{-\beta \H }~\vert 0\ra \Bigr]_{\beta, \mu \rightarrow \infty}\nonumber\\
 &=& \int d\mu ( {\cal C})_{2d} ~\Psi_0^* \Psi_0 \biggr]_{\mu \rightarrow \infty}
 \label{gen3}
 \eeqar
 As $\beta \rightarrow \infty$, only the ground state survives in the expectation value; this gives the last equality. $\Psi_0$ is the ground state wave function for $e^2 =\mu$. We need the large $e^2$ (or $\mu$) limits of $\Psi_0$ which is known from
 (\ref{c21}). Thus
 \beqar
  \int d\mu ({\cal C})_{3d} &=&\int d\mu ( {\cal C})_{2d} ~\exp\left( - {1\over 4e^2_{2d}}\int F^2\right)
  \nonumber\\
  &=& {\rm 2-dim. ~Yang\!-\!Mills ~ partition ~function ~for}~ e^2_{2d} = {\mu^2 c_A \over 2\pi}
  \nonumber\\
  &=& {\rm WZW ~partition ~function~ as} ~\mu \rightarrow \infty\nonumber\\
  &<& \infty
  \label{gen4}
  \eeqar
This leads to the (somewhat surprising) conclusion that the total volume of the configuration space is finite, even in three dimensions.
\vskip .1in\noindent{$\underline{A ~potential ~paradox ~and~ its ~resolution}$} 

We now consider a possible counter-argument for the finiteness of the total volume
of the configuration space in three dimensions. This argument is taken/adapted from
\cite{orland}, where a general analysis of many properties of the configuration space
is given.

The square of the
Euclidean distance between the gauge orbits
corresponding to the potentials $A$ and $A'$ can be defined as
\beq
L^2 (A, A') = {\rm Inf}_g \int d^3x~ \Tr ( A^g - A')^2
\label{gen5}
\eeq
The choice of the infimum over the gauge transformations $g$ picks the minimum distance between
the orbits corresponding to $A$ and $A'$. The energy functional for a configuration $A$
is given by
\beq
{\cal E}(A) = {1\over 4\mu} \int d^3x~ F^2
\label{gen6}
\eeq

Consider now the orbits of $A_i(x)$ and $A_i^{(s)}= s A_i (sx)$. It is easily checked that if $A_i(x)$ transforms as a connexion under gauge transformations, then so does $A^{(s)}$
(with a different gauge transformation matrix.) We find
\beq
L^2 (A^{(s)}, 0) = {1\over s} ~L^2 (A,0), \hskip .3in
{\cal E} (A^{(s)}) = s ~{\cal E}(A)
\label{gen7}
\eeq
As $s \rightarrow 0$, we scale up the distance of the configuration
$A$ from the trivial configuration $A=0$, yet there is no cutoff
imposed by ${\cal E}(A)$ (which scales to zero).
Thus for any configuration $A_i(x)$, we can find a sequence of configurations, parametrized by $s$,
farther and farther away with no increase in ${\cal E}$.
(Notice that this argument will not work in two spatial dimensions.)
So the question is: Since any configuration can be moved arbitrarily farther away by this scaling trick, how could one get $\int d\mu ({\cal C}) <\infty$?

The resolution of this paradox has to do with the dynamical generation of mass in three dimensions.
As we said before, integrations done with the volume measure (\ref{gen2}) can be viewed as the functional integration for a $3$-dimensional (or $(2+1)$-dimensional) Yang-Mills theory at strong coupling. In this theory there is dynamical generation of mass, so that
the effective action which controls the behavior of the integral (\ref{gen2})
has mass terms in addition to ${\cal E}(A)$. Therefore, we must
consider not just the scaling of ${\cal E}(A)$, but also of the mass term which is generated
when the functional integration is carried out.
The mass term can be seen in the Hamiltonian approach as discussed elsewhere
\cite{KN, KKN1}.
It can also be seen in a $3$-dimensional covariant approach by a resummation
technique \cite{AN, mass, JP}. For example, we may think of doing the functional integral
by progressively integrating out the higher momentum modes, obtaining a new effective action
at each stage, along the lines of the Wilsonian renormalization group.
To integrate out modes of momenta higher than some value
$M$, we rewrite the $3d$-action or energy functional as
\beq
S = {1\over \mu }\left[ {1\over 4} \int d^3x~ F^2 + M^2 S_m(A) \right]  ~- {M^2 \over \mu} S_m (A)
\label{gen8}
\eeq
Here $S_m(A)$ is a gauge-invariant mass term for the gauge potentials, the specific form of which will be briefly discussed below.
With this action, we can now consider the Feynman diagrams generated by the bracketed set of terms.
The propagators for the gauge fields are now massive and so, in integrations over the loop momenta
$k$,
 the contributions of modes
of $k \ll M$ are suppressed. The result will thus be the contribution of the Feynman
diagrams due to modes of momenta $k \gg M$. Since $S_m$ is gauge-invariant,
this gives a way of formulating the notion of the renormalization group
in a gauge-invariant way. Notice that the leading mass terms cancel out at the end, so that one is left with any mass term which is dynamically generated (plus other terms with more derivatives of the fields).
This procedure has been carried out to
one-loop order using different types of mass terms, although the interpretation there was different. For example, it was shown in
\cite{AN} that we get
\beq
S_{eff} = {1\over 4\mu} \int d^3x~ F^2 + \lambda ~S_m (A)
\label{gen9}
\eeq
where $\lambda \approx 1.2  M c_A/2\pi$. The volume element (\ref{gen2})
now becomes
\beq
 d\mu ({\cal C}, k\ll M)_{3d} 
 = {[dA]\over vol({\cal G}_*)}~  \exp \left( - {1\over 4\mu}\int F^2   - \lambda ~S_m (A)\right)\Biggr]_{\mu \rightarrow \infty}
\label{gen9a}
\eeq
The remaining integration is over modes of $A$ of momenta $k \ll M$.
Returning to the scaling of the potentials,
notice that the mass term scales as
\beq
S_m (A^{(s)}) = {1\over s}~ S_m (A)
\label{gen10}
\eeq
As $s \rightarrow 0$, we get a cutoff in the functional integral due to this mass term. This explains why it is possible to get $\int d\mu ({\cal C}) < \infty$.

\vskip .1in\noindent{$\underline{The ~nature ~of ~the ~mass~ term} $}

The qualitative nature of the result (\ref{gen9}) is not sensitive to the details of the
gauge-invariant mass term. However, for the sake of completeness, we give the 
expression for the specific mass term which was used in the calculation
of (\ref{gen9}). It is given by \cite{Nair}
\beq
S_m (A) =\int d\Omega ~K(A_n,A_{\bar n})\label{gen11}
\eeq
where $n_i$ is a (complex) three-dimensional null vector which may be parametrized as
\beq
n_i=(-\cos\theta \cos\vf-i\sin\vf,-\cos\theta
\sin\vf+i\cos\vf,\sin\theta)\label{gen12}
\eeq
In terms of this,
$A_n={1\over 2}A_in_i,~ A_{\bar n}={1\over 2} A_i{\bar n}_i$.
Further, in (\ref{gen11}), $d\Omega=\sin\theta d\theta d\vf$ and denotes integration over
the angles of $n_i$. The function $K(A_n,A_{\bar n})$ is given by
\beqar
K(A_n,A_{\bar n})&=&-{1\over\pi}\int d^2 x^T\biggl[\int d^2 z
~{\rm Tr}(A_n,A_{\bar
n})+i\pi I(A_n) + i\pi I(A_{\bar n})\biggr]\nonumber\\
I(A_n)&=&i\sum_2^\infty{(-1)^m\over m}\int {d^2 z_1 \over \pi} \ldots
{d^2z_n\over\pi}{{\rm Tr}(A_n(x_1) \ldots A_n(x_m))\over {\bar z_{12}\bar
z_{23} \dots\bar z_{m-1m}\bar z_{m1}}}\label{gen13}
\eeqar
In these expressions, $z=n\cdot{\vec x}, ~{\bar z}={\bar n}\cdot{\vec x}$ and
$x^T$ denotes the coordinate
transverse to $n_i$, i.e., ${\vec x}^T\cdot{\vec n}=0$; also
$ z_{ij}=\bar z_i-\bar z_j$.
The argument of all $A$'s in (\ref{gen13}) is the same for the transverse coordinate
$x^T$. (The complex null vectors $n$, ${\bar n}$ define a choice of complex coordinates
$n\cdot{\vec x}, ~{\bar n}\cdot{\vec x}$ at each point in space. The construction given here can thus be reinterpreted in terms of twistors for the three-dimensional space.)

If we define a complex $SL(N, \mathbb{C})$-matrix $L$ by
$n\cdot A = -n\cdot\nabla L ~L^{-1}$,
${\bar n}\cdot A = L^{\dagger -1} {\bar n}\cdot \nabla  L^\dagger$,
in a way analogous to the parametrization we used for two-dimensional Euclidean fields,
then this mass term can be written as
\beq
S_m (A) = - \int d \Omega~ dx^T~ {\cal S}_{wzw} (L^\dagger L)
\label{gen14}
\eeq
If we expand (\ref{gen13}) in powers of $A$, then the
lowest order term in $S_m$ is seen to be
\beq
S_m=~{1\over 2} \int {d^3k \over (2\pi )^3} ~
 A_i^a(-k)\biggl(\delta_{ij}-{k_ik_j\over
{\vec k}^2}\biggr)  A_j^a(k) ~ +{\cal O}(A^3)\label{gen15}
\eeq
Thus $S_m(A)$ is indeed a mass term; its gauge-invariance is evident from
(\ref{gen14}). 

It is worth emphasizing that,
for the purpose of integrating out modes of high momenta,
other mass terms, such as those given in \cite{mass, JP}, 
may also be used. Different mass terms may be viewed as different gauge-invariant completions of the basic quadratic term in (\ref{gen15}). 
As pointed out in \cite{JP}, generally, when these mass terms are used to calculate the corrections to the effective action, specifically the vacuum polarization, one gets terms which have a singularity at $k^2 =0$. In the language of unitarity cuts, when continued to
Minkowski signature, this may suggest that there are still 
massless modes. The mass term (\ref{gen14}) does not have such 
threshold singularities. This may be considered a small advantage to this particular 
mass term, but, it should be emphasized that, for the properties of the
configuration space in three Euclidean dimensions, which is what is needed for
the $(3+1)$-dimensional theory, the question of 
continuation to Minkowski signature does not arise.
\vskip .1in
We thank Abhishek Agarwal for useful comments.
This research was supported in part by the National Science Foundation 
grants PHY-0457304 and PHY-0555620 and by PSC-CUNY grants.
\fontfamily{cmr}\fontsize{11pt}{16.7pt}\selectfont


\begin{thebibliography}{99}

\bibitem{KN}D. Karabali and V.P. Nair, \NP~ {\bf B464}, 135 (1996); \PL~
{\bf B379}, 141 (1996); \IJMP~ {\bf A12}, 1161 (1997).

\bibitem{KKN1} D. Karabali, Chanju Kim and V.P. Nair, \NP~ {\bf B524}, 661 (1998).

\bibitem{KKN2} D. Karabali, Chanju Kim and V.P. Nair, \PL~ {\bf B434}, 103 (1998).

\bibitem{minic}
R.~G.~Leigh, D.~Minic and A.~Yelnikov,
\PRL~{\bf 96}, 222001 (2006) [arXiv:hep-th/0512111];
[arXiv:hep-th/0604060].
  
\bibitem{NY} V.P. Nair and A. Yelnikov, \NP~ {\bf B691}, 182 (2004).
  
\bibitem{Freidel} L. Freidel, R.G. Leigh and D. Minic, \PL~{\bf B641}, 105 (2006)
  [arXiv:hep-th/0604184]; L. Freidel, arXiv:hep-th/0604185.

 \bibitem{others} There have been a number of other analytic attempts and approaches, some of them related to ours, for Yang-Mills in 2+1 dimensions.  Some relevant articles are:\\
M.B. Halpern, \PR~ {\bf D16}, 1798 (1977); {\it ibid.}  {\bf D16}, 3515
(1977); {\it ibid.} {\bf D19}, 517 (1979);  I. Bars and F. Green, \NP~
{\bf B148}, 445 (1979); J. Greensite, \NP {\bf B158}, 469 (1979);  D.Z. Freedman and R. Khuri, \PL~ {\bf A192}, 153 (1994); M. Bauer and D.Z. Freedman, \NP~ {\bf B450}, 209 (1995); F.A. Lunev, \PL~ {\bf B295}, 99 (1992);  O. Ganor and J. Sonnenschein, \IJMP~ {\bf A11}, 5701 (1996);
S.R. Das and S. Wadia, \PR~ {\bf D53}, 5856 (1996); I.I. Kogan and
A. Kovner, \PR~ {\bf D52}, 3719 (1995); arXiv:hep-th/0205026;
P. Mansfield and D. Nolland, \JHEP~{\bf 9907}:028 (1999);
P. Mansfield, \JHEP~{\bf 0404}:059 (2004);
S.G. Rajeev, arXiv:hep-th/0401202; P. Orland, \PR~{\bf D71}, 054503 (2005);
{\it ibid.}~{\bf D74}, 085001 (2006); {\it ibid.} {\bf D75}, 025001 (2007);
arXiv:0704.0940  [hep-th].

 \bibitem{teper} M. Teper, \PR~ {\bf D59}, 014512 (1999);
B. Lucini and M. Teper, \PR~ {\bf D66}, 097502 (2002).
\bibitem{bringoltz} B.~Bringoltz and M.~Teper,
  Phys.\ Lett.\  B {\bf 645}, 383 (2007)
  [arXiv:hep-th/0611286].

\bibitem{armoni} A. Armoni and M. Shifman, \NP~{\bf B664}, 233 (2003);
\NP~{\bf B671}, 67 (2003).

\bibitem{AKN} A. Agarwal, D. Karabali and V.P. Nair,
arXiv:0705.0394.

\bibitem{orland} P. Orland, arXiv:hep-th/9607134;
\PR~{\bf D70}, 045014 (2004).

\bibitem{AN} G. Alexanian and V.P. Nair, \PL~ {\bf B352}, 435
(1995).

\bibitem{mass} W. Buchmuller and O. Philipsen, \NP~ {\bf
B443}, 47 (1995); O. Philipsen, in {\it TFT-98: Thermal Field
Theories and their Applications}, U. Heinz (ed.), hep-ph/9811469;
F. Eberlein,
\PL~ {\bf B439}, 130 (1998); \NP~ {\bf B550}, 303 (1999);
J.M. Cornwall, \PR~
{\bf D10}, 500 (1974);  {\it ibid.} {\bf D26}, 1453 (1982); \PR~
{\bf D57}, 3694 (1998).

\bibitem{JP} R. Jackiw and S-Y. Pi, \PL~ {\bf B368}, 131 (1996);
{\it ibid.} {\bf B403}, 297 (1997).

\bibitem{Nair} V.P. Nair,  \PL~ {\bf 352 B}, 117 (1995).

\end{thebibliography}
\end{document}